\documentclass[conference]{IEEEtran}
\usepackage{amsmath,graphicx}
\usepackage[affil-it]{authblk}
\usepackage{cite}
\usepackage{booktabs}
\usepackage{color}
\usepackage{url}
\usepackage{placeins}
\usepackage{amssymb}

\usepackage{stfloats}

\title{Binaural Signal Matching with Wearable Arrays for Near-Field Sources}

\author[1]{Sapir Goldring}
\author[2]{Zamir Ben Hur}
\author[2]{David Lou Alon}
\author[1]{Boaz Rafaely}

\affil[1]{School of Electrical and Computer Engineering, Ben Gurion University of the Negev}
\affil[2]{Reality Labs Research at Meta, Redmond, WA, USA}
\begin{document}

\maketitle
\begin{abstract}
Binaural reproduction methods aim to recreate an acoustic scene for a listener over headphones, offering immersive experiences in applications such as Virtual Reality (VR) and teleconferencing. Among the existing approaches, the Binaural Signal Matching (BSM) algorithm has demonstrated high quality reproduction due to its signal-independent formulation and the flexibility of unconstrained array geometry. However, this method assumes far-field sources and has not yet been investigated for near-field scenarios. This study evaluates the performance of BSM for near-field sources. Analysis of a semi-circular array around a rigid sphere, modeling head-mounted devices, show that far-field BSM performs adequately for sources up to approximately tens of centimeters from the array. However, for sources closer than this range, the binaural error increases significantly. Incorporating a near-field BSM design, which accounts for the source distance, significantly reduces the error, particularly for these very-close distances, highlighting the benefits of near-field modeling in improving reproduction accuracy.
\end{abstract}

\begin{IEEEkeywords}
binaural reproduction, near-field, virtual reality
\end{IEEEkeywords}
\section{Introduction}\label{sec:introduction}

The increasing demand for immersive auditory experiences in applications such as Virtual Reality (VR) and teleconferencing has driven significant research into binaural reproduction techniques \cite{madmoni2024design, rafaely2022spatial, richard2023audio, ben2017spectral}. Binaural reproduction aims to synthesize an acoustic scene over headphones such that it accurately represents the spatial perception of a listener \cite{madmoni2024design}.

A common method for the rendering of binaural signals uses High Order Ambisonics (HOA) signals, together with Head Related Transfer Functiong (HRTFs) \cite{poletti2005three,rafaely2010interaural}. This method has been widely studied for its ability to reproduce sound fields with high spatial accuracy \cite{rafaely2022spatial}. However, its direct application is constrained to spherical arrays, making adaptation to arbitrary microphone configurations nontrivial \cite{rafaely2022spatial}. To address this limitation, alternative approaches have been proposed, including Beamforming-Based Binaural Reproduction (BFBR) \cite{song2008using,ifergan2022selection}, parametric spatial filtering methods \cite{mccormack2022parametric}, and Binaural Signal Matching (BSM) \cite{madmoni2024design}. BSM formulates binaural reproduction as an optimization problem, minimizing the Mean-Squared Error (MSE) by matching the array steering vectors to the HRTFs, thus offering flexibility across different array geometries. 
All the methods presented share a significant limitation: they are primarily designed under the assumption of far-field plane wave propagation and are not specifically optimized for near-field conditions.

In scenarios where the source is in close proximity to the array, such as in whispering or close talk, the far-field assumption becomes inaccurate, as the sound field exhibits a spherical rather than planar wavefront \cite{fisher2010near}. This deviation may introduce errors in binaural reproduction, necessitating alternative formulations that account for near-field effects.

This study evaluates the performance of the BSM algorithm in near-field conditions.

\section{Mathematical Background}\label{sec:page_size}
This section provides mathematical background concerning array processing, binaural reproduction and near-field data. Throughout the paper, the spherical coordinate system will be used, denoted by \((r, \theta, \phi)\), where \(r\) is the radial distance from the origin, \(\theta\) is the elevation angle measured from the Cartesian \(z\)-axis downward toward the \(xy\)-plane, and \(\phi\) is the azimuth angle measured from the positive \(x\)-axis in the direction of the positive \(y\)-axis. The wave number \(k\) is defined as \(k = \frac{2\pi}{\lambda}\), where \(\lambda\) is the wavelength, and \(f\) represents the frequency. Let \(\Omega = (\theta, \phi)\) denote the direction of arrival of a sound source in the sound field.
\subsection{Array Measurement Model}

The sound field is assumed to be composed of \(Q\) sound sources, each with source signals \(s_q(k)\) assigned to the sources arriving from directions \(\{\Omega_q\}_{q=1}^Q\). This sound field is captured by an \(M\)-element microphone array centered at the origin. The recorded signals can then be represented by the following narrowband model \cite{madmoni2024design}:

\begin{equation}
    \mathbf{x}(k) = \mathbf{V}(k, \Omega)\! \mathbf{s}(k) + \mathbf{n}(k)
\label{eq:signal_model}
\end{equation}

\begin{sloppypar}
Here, \(\mathbf{x}(k) = [x_1(k), x_2(k), \dots, x_M(k)]^T\) is an \(M\)-dimensional vector containing the microphone signals. 
The matrix \(\mathbf{V}(k, \Omega) = [\mathbf{v}_1(k), \mathbf{v}_2(k), \dots, \mathbf{v}_Q(k)]\) is an \(M \times Q\) steering matrix, where each \(\mathbf{v}_q(k, \Omega_q)\) represents the steering vector corresponding to the direction of arrival \(\Omega_q\) of the \(q\)-th source. 
The vector \(\mathbf{s}(k) = [s_1(k), s_2(k), \dots, s_Q(k)]^T\) contains the source signals, and \(\mathbf{n}(k) = [n_1(k), n_2(k), \dots, n_M(k)]^T\) represents an \(M\)-dimensional additive noise vector.
\end{sloppypar}

The steering vectors can be derived analytically or numerically for specific array types  \cite{rafaely2015fundamentals}, or obtained through direct measurement. The representation of these steering vectors for different source types, including plane waves and point sources, will be demonstrated in the simulation study.

\subsection{Binaural Signal Model}

Assuming the listener's head is centered at the origin, the binaural signal at the left and right ears can be expressed as:

\begin{equation}
    p^{l/r}(k) = [\mathbf{h}^{l/r}(k)]^T \mathbf{s}(k),
    \label{eq:reproduction}
\end{equation}
where $p^{l/r}(k)$ denotes the binaural signal, $[\mathbf{h}^{l/r}(k)]^T$ is the transposed HRTF vector, and $\mathbf{s}(k)$ represents the source signal vector. The HRTF vector is given by:

\begin{equation}
    [\mathbf{h}^{l/r}(k)] = \begin{bmatrix} h^{l/r}(k, \Omega_1) & \dots & h^{l/r}(k, \Omega_Q) \end{bmatrix}^T,
\end{equation}
where $h^{l/r}(k, \Omega_q)$ is the HRTF from a source at direction $\Omega_q$. The source model (e.g., point source or plane wave) affects the complex amplitude of the HRTF \cite{kan2009psychophysical}.

\subsection{Binaural Signal Matching}

BSM estimates the binaural signal using weighted microphone signals:

\begin{equation}
    \hat{p}^{l/r}(k) = [\mathbf{c}^{l/r}(k)]^H \mathbf{x}(k),
    \label{estimated_sig}
\end{equation}
where $\mathbf{c}^{l/r}(k)$ is an $M \times 1$ complex weight vector and $(\cdot)^H$ denotes the Hermitian (conjugate transpose) operator. The filter coefficients are obtained by minimizing the MSE:

\begin{equation}
    \mathbf{c}^{l/r} = \arg \min_{\mathbf{c}} \mathbb{E} \left[ \left| p^{l/r}(k) - \hat{p}^{l/r}(k) \right|^2 \right].
\end{equation}
Under the assumption of spatially white noise and sources, and uncorrelated noise and sources, the optimal filter solution is \cite{madmoni2024design}:

\begin{equation}
    \mathbf{c}^{l/r} = \left( \mathbf{V} \mathbf{V}^H + \frac{\sigma_n^2}{\sigma_s^2} \mathbf{I}_M \right)^{-1} \mathbf{V} [\mathbf{h}^{l/r}]^*,
    \label{weights}
\end{equation}
where $\sigma_n^2$ and $\sigma_s^2$ denote the noise and signal power, respectively, and $[\mathbf{h}^{l/r}]^*$ is the complex conjugate of the HRTF vector. The normalized reproduction error is defined as:

\begin{equation}
    \epsilon^{l/r}(k) =
    \frac{ \mathbb{E}\left[ \left| p^{l/r}(k) - \hat{p}^{l/r}(k) \right|^2 \right] }
         { \mathbb{E}\left[ \left| p^{l/r}(k) \right|^2 \right] }
\end{equation}
Expanding this expression:

\begin{equation}
    \epsilon^{l/r}(k) = \frac{\sigma_s^2 \left\| \mathbf{V}^T \left(\mathbf{c}_{BSM}^{l/r}\right)^* - \mathbf{h}^{l/r} \right\|_2^2 + \sigma_n^2 \left\| \left(\mathbf{c}_{BSM}^{l/r}\right)^* \right\|_2^2}{\sigma_s^2 \left\| \mathbf{h}^{l/r} \right\|_2^2},
    \label{error}
\end{equation}
where $\|\cdot\|_2$ denotes the $L_2$-norm. This formulation quantifies the accuracy of BSM-based binaural reproduction relative to the target HRTFs.
\section{Near-Field BSM: Mathematical Framework and Modeling}

This section presents the mathematical framework for applying the BSM algorithm to near-field sources. It introduces the necessary modifications for steering functions and HRTFs to account for near-field effects, followed by an evaluation of BSM using both far-field and near-field filter designs. Finally, the modeling of near-field data is described.

\subsection{BSM Formulation for Near-Field Sources}

For a sound field composed of $Q$ point sources, the signal model given in (\ref{eq:signal_model}) can be written as:

\begin{equation}
    \mathbf{x}(k) = \mathbf{V}_{\text{nf}}(k, \Omega,r_s)\mathbf{s}(k) + \mathbf{n}(k),
\end{equation}
where $\mathbf{V}_{\text{nf}}(k, \Omega,r_s)$ is the near-field steering matrix, with $r_s$ denoting the source distance. Each column $\mathbf{v}_q(k, \Omega_q,r_q)$ represents the steering vector of a point source located at $(\Omega_q,r_q)$ relative to the $M$-element microphone array.

The true binaural signal, specified in (\ref{eq:reproduction}), becomes:

\begin{equation}
    p^{l/r}(k) = [\mathbf{h}^{l/r}_{\text{nf}}(k)]^T\mathbf{s}(k),
\end{equation}
where $\mathbf{h}^{l/r}_{\text{nf}}(k)$ is the near-field HRTF vector, incorporating the wavefront curvature and interactions with the head, torso, and pinna.

The estimated binaural signal is calculated as detailed in Equation (\ref{estimated_sig}), where the BSM filter coefficients $\mathbf{c}^{l/r}(k)$ are computed based on Equation (\ref{weights}). The difference between the far-field and near-field formulations lies in the choice of the steering matrix and HRTFs:

\begin{enumerate}
    \item \textbf{Far-Field BSM Filters (FF BSM)}: This approach assumes a plane wave model, using the far-field steering matrix and HRTFs. Here, the filter coefficients are computed using Eq. (\ref{weights}), with the substitutions:
    \begin{equation}
        \mathbf{V} = \mathbf{V}_{\text{ff}}, \quad [\mathbf{h}^{l/r}] = [\mathbf{h}^{l/r}]_{\text{ff}},
    \end{equation}
    where $\mathbf{V}_{\text{ff}}$ represents the far-field steering matrix, and $[\mathbf{h}^{l/r}]_{\text{ff}}$ are the far-field HRTFs. As a result, the computed filter coefficients for the far-field case are:
    \begin{equation}
        \mathbf{c}^{l/r}_{\text{ff}} = \left( \mathbf{V}_{\text{ff}} \mathbf{V}_{\text{ff}}^H + \frac{\sigma_n^2}{\sigma_s^2} \mathbf{I}_M \right)^{-1} \mathbf{V}_{\text{ff}} [\mathbf{h}^{l/r}]_{\text{ff}}^*.
        \label{c_ff}
    \end{equation}
    
    \item \textbf{Near-Field BSM Filters (NF BSM)}: This approach incorporates the near-field steering matrix and HRTFs, explicitly accounting for source proximity. The filter coefficients are again computed using Eq. (\ref{weights}), but with the substitutions:
    \begin{equation}
        \mathbf{V} = \mathbf{V}_{\text{nf}}, \quad [\mathbf{h}^{l/r}] = [\mathbf{h}^{l/r}]_{\text{nf}},
    \end{equation}
    As a result, the computed filter coefficients for the near-field case are:
    \begin{equation}
        \mathbf{c}^{l/r}_{\text{nf}} = \left( \mathbf{V}_{\text{nf}} \mathbf{V}_{\text{nf}}^H + \frac{\sigma_n^2}{\sigma_s^2} \mathbf{I}_M \right)^{-1} \mathbf{V}_{\text{nf}} [\mathbf{h}^{l/r}]_{\text{nf}}^*.
    \label{c_nf}
    \end{equation}
\end{enumerate}

\subsection{Modeling Near-Field Data}

Near-field BSM requires accurate steering matrices and HRTFs. However, obtaining such data experimentally is challenging due to the need for broadband sources, precise positioning, and extensive measurements \cite{brungart2002near}. To address this, a simulation-based approach is employed, leveraging the Directional Variation Function (DVF) \cite{kan2009psychophysical} and a rigid sphere model.

The DVF, defined as:

\begin{equation}
    \text{DVF} = \frac{p(r_n, \theta, \phi, k, r_a)}{p(r_f, \theta, \phi, k, r_a)}
\end{equation}
captures the pressure ratio between a near-field source at distance $r_n$ and a far-field source at distance $r_f$. The parameter $r_f$ corresponds to the reference far-field distance where the original HRTF was measured, while $r_n$ represents the near-field distance where the HRTF is estimated. The rigid sphere model, with radius $r_a$, is used to approximate the human head in acoustic simulations.
It provides an analytical solution for the acoustic pressure due to a near-field point source. This solution is expressed as \cite{fisher2010near}:

\begin{align}
    p(r_n, \theta, \phi, k, r_a) = 
    & \sum_{n=0}^{N} \sum_{m=-n}^{n} i^{-(n+1)} k h_n^{(2)}(kr_s) 4\pi i^n \nonumber \\
    & \times \left[ j_n(kr_a) - \frac{j_n^{\prime}(kr_a)}{h_n^{(2)\prime}(kr_a)} h_n^{(2)}(kr_n) \right] \nonumber \\
    & \times Y_n^m(\theta_s, \phi_s)^* Y_n^m(\theta, \phi)
    \label{eq:steering}
\end{align}
where $N$ is the Spherical Harmonics (SH) truncation order, $j_n$ and $h_n^{(2)}$ are the spherical Bessel and Hankel functions, respectively, and $Y_n^m$ are SH.

\subsubsection{Near-Field Steering Functions} 
In this work, it is assumed that the microphones are arranged in a semi-circular configuration, where all microphones are equidistant from the origin and positioned on the surface of a rigid sphere.
There was no need to use the DVF here because the near-field steering matrix in this work is computed directly using the rigid sphere model, leveraging the analytical solution in Equation (\ref{eq:steering}).

\subsubsection{Near-Field HRTFs}

Near-field HRTFs are modeled by scaling far-field HRTFs with the DVF:

\begin{equation}
    h^{l/r}(r_n, \theta, \phi, k) = \text{DVF}(r_n, r_f, \theta, \phi, k, r_a) \cdot h^{l/r}(r_f, \theta, \phi, k).
\end{equation}
Ear positions are defined at $90^\circ$ elevation, with azimuth angles of $100^\circ$ (left) and $260^\circ$ (right).

\section{Simulation Study}
This section presents a simulation study quantifying the performance of BSM 
for near-field sources. The study models near-field HRTFs and steering functions to assess reproduction accuracy using a semi-circular microphone array, reflecting potential head-mounted devices.

\subsection{Simulation Setup}
The array used is a semi-circular configuration with $M = 4$ omnidirectional microphones mounted on a rigid sphere of radius  $r = 0.1$ m, and arranged on the horizontal plane. This setup approximates head-worn arrays. The microphone positions are defined in spherical coordinates $(0.1, 90^\circ, \phi_m)$ with azimuth angles $\phi_m = [30^\circ, 80^\circ, 280^\circ, 330^\circ]$.

Steering vectors and HRTFs were computed in the SH domain, truncated at $N = 30$. Near-field HRTFs were derived from far-field HRTFs measured using the Neumann KU100 manikin from the Cologne database \cite{bernschutz2013spherical}, utilizing the DVF and the rigid sphere model  as detailed in Sec. 3.2. Simulated source distances ranged from $r_s = 0.15$ m to $r_s = 3.2$ m, with $3.2$ m representing the far-field condition.
Binaural reproduction error as in Eq. (\ref{error}) was evaluated using the two BSM filter formulations specified in Eqs. (\ref{c_ff}) and (\ref{c_nf}).

\subsection{Results}

\begin{figure}[htbp]
    \centering
    \includegraphics[width=0.48\textwidth]{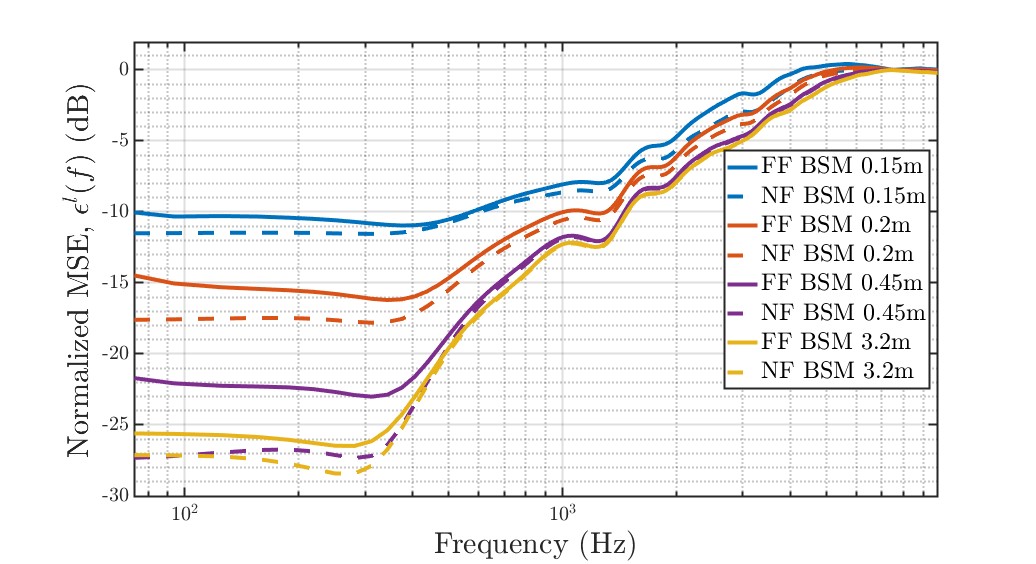}
    \caption{Normalized MSE for the left ear using far-field (solid lines) and near-field (dashed lines) BSM filters as a function of frequency for source distances from 0.15 to 3.2 meters.}
    \label{fig:Mixed_MSE}
\end{figure}

Figure \ref{fig:Mixed_MSE} presents the normalized MSE for the left ear as defined in Equation (\ref{error}) for both far-field and near-field BSM filters, computed using Equations (\ref{c_ff}) and (\ref{c_nf}), respectively. Dashed lines indicate near-field filter errors, while solid lines represent far-field filter errors.

The errors are shown for frequencies between \(75 \, \text{Hz}\) and \(10\, \text{kHz}\) and the results are presented for several source distances, with \(3.2 \, \text{m}\) representing the far-field distance at which the original far-field HRTFs were measured. As expected, the error above 2 kHz is large for all cases due to the increasing SH order of the HRTFs, as noted in \cite{madmoni2024design}. Below $2$ kHz, the binaural error for FF BSM is higher than for NF BSM. This is expected, as the NF BSM incorporates source distance information, allowing for improved accuracy in near-field conditions. 
At closer distances, particularly at $r_s = 0.15 \text{ m} \text{ and } 0.2 \text{ m}$
, corresponding to a source of only $5-10 \, \text{cm}$ from the rigid sphere surface ($r = 0.10 \, \text{m}$), a significant increase in error is observed. It appears that at such close distances, the performance of NF BSM is limited.

\section{Discussion and Conclusions}
This paper investigated the development of a framework for performing near-field binaural reproduction, including the signal model and a proposed method for computing near-field BSM that incorporates source distance information. The approach was developed using simulated near-field data, based on previous work, and adapted for binaural reproduction in near-field conditions.

The results show that for sources at moderate distances from the array, far-field BSM provides reasonable accuracy. However, for closer sources, the near-field BSM approach significantly reduces the MSE, demonstrating the advantage of incorporating source distance information. Nevertheless, for very close distances, particularly when the source is within a few centimeters of the array, the binaural error remains relatively high, even when using near-field BSM. This suggests that additional factors affect the performance at extreme near-field distances. Future work should investigate the underlying reasons for this limitation and explore potential solutions. Additionally, a listening test should be conducted to assess the perceptual impact of these errors and validate the objective measures used in this study.
% References


%\bibliographystyle{IEEEtran.bst}
%\bibliography{refs}
% ==== Begin bibliography ====

\end{document}